\documentclass[twocolumn,showpacs,preprintnumbers,floatfix,
               amsmath,amssymb]{revtex4}
\usepackage{longtable}
\usepackage{graphics}
\usepackage{epsfig}
\def\pmbf#1{{\mbox{\boldmath${#1}$}}}

\begin{document}

\title{Neutrino-pasta scattering: the opacity of nonuniform 
       neutron-rich matter}

\author{C.J. Horowitz}\email{horowit@indiana.edu} 
\author{M.A. P\'{e}rez-Garc\'{\i }a}\email{mperezga@indiana.edu}
\affiliation{Nuclear Theory Center and Department of Physics, 
             Indiana University,  Bloomington, Indiana 47405}
\author{J. Piekarewicz}\email{jorgep@csit.fsu.edu}
\affiliation{Department of Physics,
         Florida State University, Tallahassee, FL\ \ 32306}


\date{\today}

\begin{abstract}
Neutron-rich matter at subnuclear densities may involve complex
structures displaying a variety of shapes, such as spherical,
slablike, and/or rodlike shapes. These phases of the {\it nuclear
pasta} are expected to exist in the crust of neutron stars and in
core-collapse supernovae. The dynamics of core-collapse supernovae is
very sensitive to the interactions between neutrinos and
nucleons/nuclei. Indeed, neutrino excitation of the low-energy modes
of the pasta may allow for a significant energy transfer to the
nuclear medium, thereby reviving the stalled supernovae shock. The
linear response of the nuclear pasta to neutrinos is modeled via a
simple semi-classical simulation. The transport mean-free path for
$\mu$ and $\tau$ neutrinos (and antineutrinos) is expressed in terms
of the static structure factor of the pasta, which is evaluated
using Metropolis Monte Carlo simulations.
\end{abstract}
\maketitle
%
\section{Introduction}
\label{intro}

Neutron-rich matter may have a complex structure at densities just
below that of normal nuclei. This is because all conventional matter
is {\it frustrated}. While nucleons are correlated at short distances
by attractive strong interactions, they are anti-correlated at large
distances because of the Coulomb repulsion. Often these short and
large distance scales are well separated, so nucleons bind into nuclei
that are segregated in a crystal lattice. However, at densities of
the order of $10^{13}\!-\!10^{14}$~g/cm$^3$ these length scales are
comparable~\cite{Rav83_PRL50}. Competition among these interactions
({\it i.e., frustration}) becomes responsible for the development of 
complex structures with many possible nuclear shapes, such as spheres,
cylinders, flat plates, as well as spherical and cylindrical
voids~\cite{Has84_PTP71}. The term {\it pasta phases} has been coined
to describe these complex structures~\cite{Rav83_PRL50}, and many
calculations of their ground-state structure have already been
reported~\cite{Rav83_PRL50,Has84_PTP71,Wil85_NPA435,Lor93_PRL70,
Oya93_NPA561,Sum95_NPA595}. While the study of these pasta phases is
interesting in its own right, it becomes even more so due to its
relevance to the structure of the inner crust of neutron stars and to
the dynamics of core-collapse supernovae.

Frustration, a phenomenon characterized by the existence of a very
large number of low-energy configurations, emerges from the
impossibility to simultaneously minimize all ``elementary''
interactions. Should a proton in the pasta join a nuclear cluster to
benefit from the nuclear attraction or should it remain well separated
to minimize the Coulomb repulsion?  Frustration, a term that appears
to have been coined in the late seventies~\cite{Tou77_CP2,Vil77_JP10},
is prevalent in complex systems ranging from
magnetism~\cite{Lie86_SV251,Die94_WS} to protein
folding~\cite{Cam96_PRL77}. In condensed-matter systems, frustration
dates back to the 1950 study of Ising antiferromagnets on triangular
lattices by Wannier~\cite{Wan50_PR79}. Three antiferromagnetically
coupled spins fixed to the sites of an equilateral triangle cannot
minimize all interactions simultaneously: once two spins are
anti-aligned, the third one cannot be antiparallel to both of them.
Further, in Ref.~\cite{Kir78_PRB17} it has been shown that finding 
the true ground state---among the many metastable states---of a spin 
glass shares features in common with {\it NP-complete} problems, such 
as the {\it traveling salesman problem} of fame in the theory of
combinatorial optimization~\cite{Pap98_Dov}. Finally, because of the 
preponderance of low-energy states, frustrated systems display unusual 
low-energy dynamics. 

Because of the complexity of the system, almost no work has been done 
on the low-energy dynamics of the pasta or on its response to 
weakly-interacting probes. In this paper we study the excitations of
the pasta via a simple semi-classical simulation similar to those used
to describe heavy-ion collisions. Heavy-ion collisions can produce
hot, dense matter. However, by carefully heating the system and
allowing it to expand, heavy-ion collisions can also study matter at
low densities. Multi-fragmentation, the breakup of a heavy ion into
several large fragments, shares many features with pasta formation, as
they are both driven by the same volume, surface, and Coulomb
energies. There have been several classical
\cite{Bod80_PRC22,Sch87_PRC36} and quantum-molecular-dynamics
(QMD)~\cite{Pei91_PLB260} simulations of heavy-ion collisions.  These
same approaches may be applied to the nuclear pasta by employing a
simulation volume and periodic boundary conditions. One great
advantage of such simulations is that one can study pasta formation in
an unbiased way without having to assume particular shapes or
configurations from the outset. While QMD has been used before to
study the structure of the
pasta~\cite{Mar98_PRC57,Wat03_PRC68,Wat03_xxx}, no calculations of its
linear response to weakly-interacting probes ({\it e.g.,} neutrinos) 
have been reported.

Neutrino interactions are crucial in the dynamics of core-collapse
supernovae because neutrinos carry 99\% of the energy. Neutrinos are
initially trapped due to the large coherent neutrino-nucleus elastic
scattering cross section. This trapping is important for the
electron-per-baryon fraction $Y_e$ of the supernova core, as it
hinders any further conversion of electrons into neutrinos.  However,
with increasing density Coulomb interactions between ions lead to
significant ion screening of neutrino-nucleus elastic
scattering~\cite{Hor97_PRD55}.  As the density is increased still
further, the ions react to form nuclear pasta and one needs to
calculate neutrino-pasta interactions.

In the pasta phase one can have coherent neutrino scattering from
density contrasts, such as {\it Swiss cheese} like voids. Critical
fluctuations could significantly increase the cross section, thereby
greatly reducing the neutrino mean-free path.  As the existence of
many low-energy configurations is the benchmark of frustrated systems,
one expects large configuration mixing among the various different
pasta shapes. This mixing often leads to interesting low-energy
collective excitations. As it will be shown later (see
Fig.~\ref{Fig2}) at subnuclear densities of the order of
$10^{13}$~g/cm$^3$, the pasta resembles a collection of spherical
neutron-rich nuclei embedded in a dilute neutron gas. Neutron-rich
nuclei with large neutron skins have {\it Pygmy giant resonances},
involving collective oscillations of the neutron skin against the
symmetric core~\cite{Iga86_NPA457,Vre01_PRC63}. We expect that the
soft neutron-rich pasta will have many low-energy collective
oscillations.  This could provide important physics that is presently
missing from core-collapse supernovae simulations. Neutrino excitation
of the low-energy pasta modes may allow for a significant energy
transfer to the nuclear medium, potentially reviving the stalled
supernovae shock.  To our knowledge, there have been no calculations
of these effects.  Note, however, that Reddy, Bertsch, and
Prakash~\cite{Red00_PLB475} have found that coherent neutrino
scattering from a nonuniform kaon condensed phase greatly decreases
the neutrino mean free path.

Present models of the equation of state for supernovae simulations,
such as that of Lattimer and Swesty~\cite{Lat92_NPA535}, describe the
system as a liquid drop for a single representative heavy
nucleus surrounded by free alpha particles, protons, and neutrons.
One then calculates neutrino scattering from these constituents---by
arbitrarily matching to a high-density uniform phase~\cite{Lie02_APJ}.  
Unfortunately, this approximation is uncontrolled as it neglects many 
important interactions between nuclei. By simulating the pasta phase 
directly in the nucleon coordinates, one hopes to improve on this 
matching and to understand its limitations.

There is a duality between microscopic descriptions of the system
in terms of nucleon coordinates and ``macroscopic'' descriptions 
in terms of effective nuclear degrees of freedom. Thus, a relevant 
question to pose is as follows: when does a neutrino scatter from 
a nucleus and when does it scatter from an individual nucleon? At 
the Jefferson Laboratory a similar question is studied; when,
{\it i.e.,} at what momentum transfer, does a photon couple to a 
full hadron and when to an individual quark? Models of the
quark/hadron duality have provided insight on how descriptions in 
terms of hadron degrees of freedom can be equivalent to descriptions 
in terms of quark coordinates~\cite{Jes02_PRD65}. Here we are
interested in nucleon/nuclear duality, that is, how can nuclear 
models incorporate the main features of microscopic nucleon 
descriptions?

The manuscript has been organized as follows. In Sec.~\ref{formalism}
the semi-classical formalism is introduced. A very simple (perhaps
minimal) model is employed that contains the essential physics of
frustration. The linear response of the pasta to neutrino scattering,
in the form of a static structure factor, is discussed in
Sec.~\ref{scattering}. Results are presented in Sec.~\ref{results},
while conclusions and future directions are reserved to
Sec. \ref{conclusions}.

\section{Formalism}
\label{formalism}

In this section we introduce a classical model that while simple,
contains the essential physics of frustration. That is, competing
interactions consisting of a short-range nuclear attraction and a
long-range Coulomb repulsion.  We model a charge-neutral system of
electrons, protons, and neutrons.  The electrons are assumed to be a
degenerate free Fermi gas of density $\rho_e\!=\!\rho_p$ and the
nucleons interact via a classical potential. The only quantum aspects
of the calculation are the use of an effective temperature and
effective interactions to simulate effects associated with quantum
zero-point motion. Of course more elaborate models are possible and
these will be presented in future contributions.  For these first
simulations we adopted a very simple version that displays the
essential physics of nucleons clustering into pasta in a transparent
form. Moreover, this simple model facilitates simulations with a
relatively large numbers of particles, a feature that is essential to
estimate and control finite-size effects.

The total potential $V_{\rm tot}$ energy is assumed to be a sum over 
two-body interactions $V_{ij}$ of the following form:
\begin{equation}
 V_{tot}=\sum_{i<j} V(i,j) \;,
 \label{vtot}
\end{equation}
where the ``elementary'' two-body interaction is given by
\begin{equation}
 V(i,j) = a e^{-r_{ij}^{2}/\Lambda} +
       \Big[b+c\tau_z(i)\tau_z(j)\Big]
	e^{-r_{ij}^{2}/2\Lambda}+V_{\rm c}(i,j)\;.
 \label{v}
\end{equation}
Here the distance between the particles is denoted by 
$r_{ij}=|{\bf r}_i\!-\!{\bf r}_j|$ and the isospin of the 
$j_{\rm th}$ particle is $\tau_z(j)\!=\!1$ for a proton and
$\tau_z(j)\!=\!-\!1$ for a neutron.  The model parameters $a$, $b$,
$c$, and $\Lambda$ will be discussed below. Suffices to say that the
above interaction includes the characteristic intermediate-range
attraction and short-range repulsion of the nucleon-nucleon ($NN$)
force. Further, the isospin dependence of the potential insures that
while pure neutron matter is unbound, symmetric nuclear matter is
bound appropriately. Finally, a screened Coulomb interaction of the
following form is included:
\begin{equation}
  V_{\rm c}(i,j)=\frac{e^{2}}{r_{ij}}e^{-r_{ij}/\lambda}
                 \tau_p(i)\tau_p(j) \;,
 \label{vc}
\end{equation}
where $\tau_p(j)\!=\!(1\!+\!\tau_z(j))/2$ and $\lambda$ is the 
screening length that results from the slight polarization of 
the electron gas. That is, the relativistic Thomas-Fermi screening 
length is given by
\begin{equation}
 \lambda=\frac{\pi}{e} 
 \left(k_{\rm F}\sqrt{k_{\rm F}^2+m_e^2}\right)^{-1/2}
 \hspace{-0.2cm}\;,
 \label{lambda}
\end{equation}
Note that the electron Fermi momentum has been defined by 
$k_{\rm F}\!=\!(3\pi^2\rho_e)^{1/3}$ and $m_e$ is the electron
mass~\cite{Fet71_MH,Chi77_AOP108}. Unfortunately, while the screening
length $\lambda$ defined above is smaller than the length $L$ of our
simulation box, it is not significantly smaller (unless a
prohibitively large number of particles is used). Therefore, to
control finite-size effects we were forced to arbitrarily decrease the
value of $\lambda$ (see Sec.~\ref{results}).

The simulations are carried out in a canonical ensemble with a fixed 
number of particles $A$ at a temperature $T$. The volume $V$ at a
fixed baryon density $\rho$ is simply $V\!=\!A/\rho$. To minimize 
finite-size effects we use periodic boundary conditions, so that the 
distance $r_{ij}$ is calculated from the $x$, $y$, and $z$ coordinates 
of the $i_{\rm th}$ and $j_{\rm th}$ particles as follows:
\begin{equation}
 r_{ij}=\sqrt{[x_i-x_j]^2 + [y_i-y_j]^2 + [z_i-z_j]^2} \;,
 \label{pbc}
\end{equation}
where the periodic distance, for a cubic box of side $L=V^{1/3}$,
is given by 
\begin{equation}
 [l]={\rm Min}( |l|, L-|l| ) \;.
\end{equation}

The potential energy defined in Eq.~(\ref{vtot}) is independent of
momentum. Therefore, the partition function for the system factors
into a product of a partition function in momentum space---that plays
no role in the computation of momentum-independent observables---times
a coordinate-space partition function of the form
\begin{equation}
 Z(A,T,V)=\int d^3r_{1} \cdots d^3r_{A}
 \exp\left(-V_{\rm tot}/T\right)\;.
 \label{zv}
\end{equation}
Note that the 3-dimensional integrals are performed over the
simulation volume $V$.

The average energy of the system 
$\langle E \rangle\!=\!\langle K \rangle\!+\!\langle V_{\rm tot} 
\rangle$ is made of kinetic ($K$) and 
potential ($V_{\rm tot}$) energy contributions. As the
(momentum-independent) interactions have no impact on
momentum-dependent quantities, the expectation value of the 
kinetic energy reduces to its classical value, that is,
\begin{equation}
 \langle K \rangle=\frac{3}{2} A T \;.
\end{equation}
In turn, the expectation value of the potential energy may be
computed from the coordinate-space partition function as
follows:
\begin{equation}
 \langle V_{\rm tot}\rangle= \frac{1}{Z(A,T,V)}
                \int d^3r_{1}\cdots d^3r_{A} 
                V_{\rm tot}\exp\left(-V_{\rm tot}/T\right)\;.
 \label{Vtot2}
\end{equation}
In summary, a classical system has been constructed with a 
total potential energy given as a sum of two-body 
momentum-independent interactions [see Eq.~(\ref{v})]. Any 
momentum-independent observable of interest can be calculated 
from the partition function [Eq.~(\ref{zv})], which we evaluate 
via Metropolis Monte-Carlo integration~\cite{Met53_JCP21}.


\begin{figure}[ht]
\begin{center}
\includegraphics[width=3.5in,angle=0,clip=false]{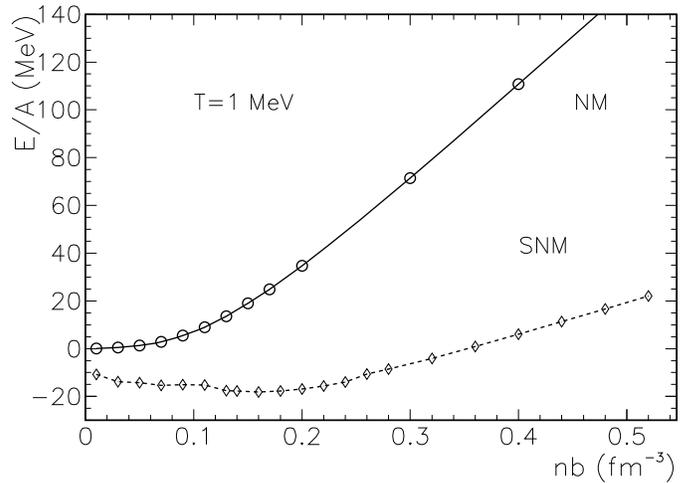}

\caption{Energy per particle for symmetric (dashed line) and for pure-neutron matter (solid) versus baryon density $nb$ at a temperature of $T\!=\!1$ MeV.}
\label{Fig1}
\end{center}
\end{figure}


We now return to discuss the choice of model parameters. The
constants $a$, $b$, $c$, and $\Lambda$ in the two-body interaction
Eq.~(\ref{v}) were adjusted---approximately---to reproduce the
following bulk properties: a) the saturation density and binding
energy per nucleon of symmetric nuclear matter, b) (a reasonable value
for) the binding energy per nucleon of neutron matter at saturation
density, and c) (approximate values for the) binding energy of a 
few selected finite nuclei. The temperature was arbitrarily fixed
at 1 MeV for all the calculations. Note that the parameter set 
employed in these calculations (and displayed in Table~\ref{Table1}) 
has yet to be carefully optimized. We reiterate that for these first 
set of simulations, the interaction is sufficiently accurate to
describe the essential physics of the pasta. Indeed, this is
illustrated in Fig.~\ref{Fig1} and Table~\ref{Table2}. In
Fig.~\ref{Fig1} the average potential energy versus density
at a temperature of $T\!=\!1$ MeV is displayed for a simulation 
of symmetric nuclear matter containing $A\!=\!400$ particles and,
as is customary, assuming no Coulomb interactions. Also shown
in the figure is the potential energy for pure neutron matter 
calculated with $N\!=\!200$ particles. In the case of finite
nuclei (also calculated at $T\!=\!1$ MeV) the full Coulomb 
interaction is included using a screening length $\lambda$ 
much larger than the resulting root-mean-square radius of the 
nucleus. Simulations based on a Metropolis Monte-Carlo 
algorithm were used to compute the average potential energy,
starting with nucleons distributed uniformly in a sphere 
with a radius comparable to the expected size of the 
nucleus; this sphere was placed in the center of a very large 
box. Results of the simulations and comparison with experimental
values have been collected in Table~\ref{Table2}. Note that the
simulation results are for the potential energy only. If the 
kinetic energy per nucleon ($3T/2$) is added to these values, 
the nuclei in Table~\ref{Table2} would be slightly underbound.  Furthermore, finite nuclei are only metastable in this semiclassical approximation.  Nucleons can evaporate over a very long time scale.  However, this is not expected to significantly impact the pasta phases since these already have free nucleons. 

\begin{table}
\caption{Model parameters used in the calculations.}
 \begin{ruledtabular}
 \begin{tabular}{cccc}
   $a$     & $b$     & $c$    & $\Lambda$ \\
   \hline
   110 MeV & -26 MeV & 24 MeV & 1.25 fm$^2$
 \label{Table1}
 \end{tabular}
\end{ruledtabular}
\end{table}

\begin{table}
\caption{Binding energy per nucleon for various closed
         shell nuclei. Note that all energies are in MeV
         and that the Monte-Carlo results include only
         $\langle V_{\rm tot} \rangle$.}
 \begin{ruledtabular}
 \begin{tabular}{cccc}
  Nucleus & Monte Carlo $\langle V_{\rm tot} \rangle$ & Experiment \\
  \hline
  $^{16}$O   &  $-7.56 \pm 0.01$ & $-7.98$ \\
  $^{40}$Ca  &  $-8.75 \pm 0.03$ & $-8.45$ \\
  $^{90}$Zr  &  $-9.13 \pm 0.03$ & $-8.66$ \\
  $^{208}$Pb &  $-8.2 \pm 0.1$ & $-8.45$
 \label{Table2}
 \end{tabular}
\end{ruledtabular}
\end{table}

\section{Neutrino Scattering}
\label{scattering}

The model is used to describe neutrino scattering from nonuniform
neutron-rich matter. As neutrino interactions play a predominant
role in core-collapse supernovae, one is interested in
understanding how the neutrinos diffuse and how do they exchange 
energy. In this first paper we focus on the transport mean free path 
for $\nu_\mu$ and $\nu_\tau$, which lack charged-current interactions 
at low energies. Their mean-free path is dominated by neutral current 
neutrino-nucleon scattering.

The free-space cross section for neutrino-nucleon elastic scattering 
is given by
\begin{equation}
 \frac{d\sigma}{d\Omega} = \frac{G_{F}^2 E_\nu^2}{4\pi^2} 
 \left[c_a^2 (3-\cos\theta) + 
       c_v^2 (1+\cos\theta)\right] \;,
 \label{cross}
\end{equation}
where $G_{F}$ is the Fermi coupling constant, $E_\nu$ the neutrino 
energy and $\theta$ the scattering angle. Note that this equation 
neglects weak magnetism and other corrections of order $E_\nu/M$, 
with $M$ the nucleon mass~\cite{Hor02_PRD65}.

In the absence of weak magnetism, the weak neutral current $J_\mu$ 
of a nucleon has axial-vector ($\gamma_5\gamma_\mu$) and vector 
$\gamma_\mu$ contributions. That is,
\begin{equation}
  J_\mu=c_a\gamma_5\gamma_\mu + c_v\gamma_\mu \;.
\end{equation}
The axial coupling constant is,
\begin{equation}
 c_a=\pm \frac{g_a}{2} \quad (g_a=1.26)\;,
 \label{ca}
\end{equation}
with the $+$ sign for neutrino-proton and the $-$ sign for 
neutrino-neutron scattering. The weak charge of a proton 
$c_v$ is suppressed by the weak-mixing (or Weinberg) angle
$\sin^{2}\theta_{\rm W}\!=\!0.231$, 

\begin{equation}
  c_v=\frac{1}{2}-2\sin^{2}\theta_{\rm W}
     =0.038 \approx 0\;.
 \label{cvp}
\end{equation}
In contrast, the weak charge of a neutron is both large
and insensitive to the weak-mixing angle: $c_v\!=\!-\!1/2$.
The transport mean-free path $\lambda_t$ is inversely proportional
to the transport cross section $\sigma_t$ and is given by the
following expression:
\begin{equation}
 \sigma_t = \int d\Omega \frac{d\sigma}{d\Omega}(1-\cos\theta)
          = \frac{2G_{F}^2 E_\nu^2}{3\pi}\left(5c_a^2+c_v^2\right)\;. 
 \label{sigmat}
\end{equation}
The weighting factor ($1\!-\!\cos\theta$) included in the definition 
of the transport cross section favors large-angle scattering, as
momentum is transferred more efficiently into the medium. As a result, 
the axial-vector contribution $c_a^2$ dominates the cross section.  
Assuming that the scattering in the medium is the same as in free 
space, the transport mean-free path becomes 
\begin{equation}
 \lambda_t=\left(\rho_p\sigma_t^p+\rho_n\sigma_t^n\right)^{-1}\;.
 \label{lambdat}
\end{equation}
Here $\rho_p$($\rho_n$) is the proton(neutron) density and 
$\sigma_t^p$($\sigma_t^n$) is the transport cross section 
for scattering from a proton(neutron).

If nucleons cluster tightly into nuclei or into pasta, then the
scattering from different nucleons could be coherent. This will
significantly enhance the cross section as it would be
proportional to the {\it square} of the number of nucleons 
\cite{Fre77_ARNS27}. The contribution from the vector current 
is expected to be coherent. Instead, the strong spin and isospin 
dependence of the axial current should reduce its coherence.
This is because in nuclei---and presumably in the pasta---most 
nucleons pair off into spin singlet states (note that in the 
nonrelativistic limit the nucleon axial-vector current becomes
$\gamma_5{\pmbf\gamma}\tau_{z} \rightarrow {-\pmbf\sigma}\tau_{z}$).
Therefore, in this paper we focus exclusively on coherence effects 
for the vector current.

Coherence is important in x-ray scattering from crystals. Because
the x-ray wavelength is comparable to the inter-particle spacing, 
one needs to calculate the relative phase for scattering from 
different atomic planes and then sum over all planes. 
Neutrino-pasta scattering involves a similar sum because the neutrino 
wavelength is comparable to the inter-particle spacing and even to 
the inter-cluster spacing. Therefore, one must calculate the relative 
phase for neutrino scattering from different nucleons and then add
their contribution coherently. This procedure is embodied in the
static structure factor $S(q)$.

The dynamic response of the system to a probe of momentum transfer 
${\bf q}$ and energy transfer $\omega\!>\!0$ that couples to the 
weak charge density $\hat{\rho}({\bf q})$ is given 
by~\cite{Fet71_MH}
\begin{equation}
 S({\bf q},\omega)=\sum_{n\neq 0}
 \Big|\langle\Psi_{n}|\hat{\rho}({\bf q})|\Psi_{0}\rangle\Big|^{2}
 \delta(\omega-\omega_{n}) \;,
 \label{Sofqw}
\end{equation}
where $\omega_{n}$ is the energy difference between the excited 
state $|\Psi_{n}\rangle$ and the ground state $|\Psi_{0}\rangle$.
In linear response theory, namely, assuming that the process can
be treated in lowest order (an excellent approximation for
neutrino scattering) the cross section can be directly related 
to the dynamic response of the system. In the case that the 
individual excited states may not be resolved, then one integrates
over the energy transfer $\omega$ to obtain the static structure
factor. Here we define the static structure factor {\it per neutron}
as follows:
\begin{equation}
  S({\bf q})=\frac{1}{N}
             \int_{0}^{\infty}S({\bf q},\omega)d\omega
            =\frac{1}{N}
             \sum_{n\neq 0}\Big|\langle\Psi_{n}|\hat{\rho}
             ({\bf q})|\Psi_{0}\rangle\Big|^{2} \;,
 \label{sq}
\end{equation}
with the weak vector-charge density given by
\begin{equation}
 {\rho}({\bf q})=\sum_{i=1}^{N} 
  \exp(i{\bf q}\cdot{\bf r}_i) \;,
 \label{rhoq}
\end{equation}
where the sum in Eq. (\ref{rhoq}) is only over neutrons.

The cross section {\it per neutron} for neutrino 
scattering from the whole system is now given by
\begin{equation}
 \frac{1}{N}\frac{d\sigma}{d\Omega}=S({\bf q}) 
 \frac{G_{F}^2 E_\nu^2}{4\pi^2}\frac{1}{4}(1+\cos\theta)\;.
 \label{sigma2}
\end{equation}
Note that the weak charge of the nucleon ($c_v\!\approx\!0$ for
protons and $c_v\!=\!-\!1/2$ for neutrons) has been incorporated into
the above cross section, so that the normalization of the weak
vector-charge density is ${\rho}(q\!=\!0)\!=\!N$. Further,
Eq.~(\ref{sigma2}) is the cross section per neutron obtained from
Eq.~(\ref{cross}) (with $c_a\!=\!0$) multiplied by S({\bf q}). 
This indicates that S({\bf q}) embodies the effects from
coherence. Finally, note that the momentum transfer is related to the
scattering angle through the following equation:
\begin{equation}
 q^2=2E_\nu^2(1-\cos\theta) \;.
 \label{q}
\end{equation}
Two assumptions have been made in the derivation of
Eq.~(\ref{sigma2}).  First, no contribution from the axial current to
the cross section has been included, because nucleons pair into
spin-zero states. Second, the excitation energy transferred to the
nucleons is small and we have summed over all possible excitation
energies.

The static structure factor has important limits. A particularly
useful form in which to discuss them invokes completeness on 
Eq.~(\ref{sq}). That is,
\begin{equation}
 S({\bf q})= \frac{1}{N}
             \left(\langle\Psi_{0}|\hat{\rho}^{\dagger}({\bf q})
             \hat{\rho}({\bf q})|\Psi_{0}\rangle
            -\Big|\langle\Psi_{0}|\hat{\rho}
             ({\bf q})|\Psi_{0}\rangle\Big|^{2}\right)\;.
 \label{sq2}
\end{equation}
The last term in the above expression represents the elastic form 
factor of the system, which only contributes at $q\!=\!0$.
In the limit of $q\!\rightarrow\!0$, the weak charge density
[Eq.~(\ref{rhoq})] becomes the number operator for neutrons
$\hat{\rho}(q\!=\!0)\!=\!\hat{N}$, so that the static structure
factor reduces to,
\begin{equation}
 S(q=0)=\frac{1}{N}\left(\langle\hat{N}^{2}\rangle-
                         \langle\hat{N}\rangle^{2}\right).
 \label{S0}
\end{equation}
Thus, the $q\!\rightarrow\!0$ limit of the static structure factor is
related to the fluctuations in the number of particles, or
equivalently, to the density fluctuations. These fluctuations are,
themselves, related to the compressibility and diverge at the
critical point \cite{liquid}.  To discuss the large 
$q$ limit, Eq.~(\ref{rhoq}) is substituted into Eq.~(\ref{sq2}) 
to yield
\begin{equation}
 S({\bf q})\!=\! \frac{1}{N}\left(\sum_{i,j}^{N}
   \langle\Psi_{0}|\exp(i{\bf q}\cdot{\bf r}_{ij})
   |\Psi_{0}\rangle \!-\! 
   \Big|\langle\Psi_{0}|\hat{\rho}({\bf q})
   |\Psi_{0}\rangle\Big|^{2}\right),
 \label{sq3}
\end{equation}
In the $q\rightarrow \infty$ limit, all the terms in the sum 
with $i\!\neq\!j$, as well as the second term in the above
expression, oscillate to zero. This only leaves the $i\!=\!j$ 
terms, which there are $N$ of them so that
\begin{equation}
 S(q\rightarrow\infty)=1 \;.
 \label{slarge}
\end{equation}
This result indicates that if the neutrino wavelength is much shorter 
than the inter-particle separation, the neutrino only resolves one
nucleon at a time. This corresponds to quasielastic scattering where 
the cross section per nucleon in the medium is the same as in free 
space.

One can calculate the static structure factor from the neutron-neutron 
correlation function which is defined as follows:
\begin{equation}
 g({\bf r})=\frac{1}{N\rho_n} \sum_{i\neq j}^{N}
      \langle\Psi_{0}|\delta({\bf r}- {\bf r}_{ij})
      |\Psi_{0}\rangle \;.
 \label{gr}
\end{equation}
The two-neutron correlation function ``asks'' (and ``answers'') the 
following question: if one ``sits'' on a neutron, what is the 
probability of finding another one a distance $|{\bf r}|$ away. 
The correlation function is normalized to one at large distances 
$g(r\rightarrow\infty)\!=\!1$; this corresponds to the average 
density of the medium. The static structure factor is obtained
from the Fourier transform of the two-neutron correlation function.
Comparing with Eq. (\ref{sq3}) this yields,
\begin{equation}
 S({\bf q})=1+\rho_n\int d^3r \Big(g({\bf r})-1\Big) 
            \exp(i{\bf q}\cdot {\bf r}) \;.
 \label{sq4}
\end{equation}
The $i\!=\!j$ terms in Eq.~(\ref{sq3}) gives the leading $1$ in the
above expression, while the elastic form factor 
$|\langle\Psi_{0}|\hat{\rho}({\bf q})|\Psi_{0}\rangle|^{2}$ yields
the $-1$ in the integrand of Eq.~(\ref{sq4}).

To obtain the transport cross section we proceed, as in 
Eq.~(\ref{cross}), to integrate the angular-weighted cross section 
$d\sigma/d\Omega (1-\cos\theta)$ over all angles. That is,
\begin{equation}
 \sigma_t=\frac{1}{N}\int d\Omega\frac{d\sigma}{d\Omega}
 (1-\cos\theta)=\langle S(E_\nu)\rangle\sigma_t^0.
\label{sigmat2}
\end{equation}
Note that the free neutron cross $\sigma_t^0$ follows directly 
from Eq.~(\ref{sigmat}) in the limit of $c_{a}\!\equiv\!0$, 
\begin{equation}
 \sigma_t^0 = \frac{G_{F}^2 E_\nu^2}{6\pi} \;.
 \label{sigmat0}
\end{equation}
Further, the angle averaged static structure factor has been
defined as follows:
\begin{equation}
 \langle S(E_\nu)\rangle\equiv\frac{3}{4}
 \int_{-1}^1 dx (1-x^2)S\bigl(q(x,E_\nu)\bigr) \;,
 \label{save}
\end{equation}
where the static structure factor $S(q)$ depends on neutrino energy 
and angle through Eq.~(\ref{q}), that is, $q^2\!=\!2E_\nu^2(1-x)$ 
with $x\!=\!\cos\theta$. Using Eq.~(\ref{sq4}) and switching the 
orders of integration, this can be written as
\begin{equation}
 \langle S(E_\nu)\rangle = 1+\frac{4\pi\rho_n}{E_\nu^2}
 \int_0^\infty dr f(2E_\nu r)\Big(g(r)-1\Big) \;. 
\label{save2}
\end{equation}
Note that the following function has been introduced
\begin{equation}
  f(t)=72(\cos t + t \sin t -1)/t^4 
      -6(5\cos t + t \sin t +1)/t^2 \;.
 \label{fave}
\end{equation}
Finally, the transport mean free path ($\lambda_t=1/\sigma_t\rho_n$) 
is given by
\begin{equation}
 \lambda_t^{-1}=\sigma_t^0\rho_n \langle S(E_\nu) \rangle \;.
\label{mfp}
\end{equation}
In this way $\langle S(E_\nu) \rangle$ describes how coherence 
modifies the mean-free path. In the next section, simulation results 
for the two-neutron correlation function, the static structure factor,
and the angle averaged static structure factor will be presented.

\section{Results}
\label{results}

In this section we present our simulation results. As an example of
a typical low-density condition we consider a subnuclear density of
$\rho\!=\!0.01~{\rm fm}^{-3}$ (about $1/16$ of normal nuclear
density), a temperature $T\!=\!1$ MeV, and an electron fraction of
$Y_e\!=\!0.2$.  Proto-neutron stars have electron fractions that
start out near $1/2$ and drop with time, so $Y_e\!=\!0.2$ represents 
a typical neutron-rich condition. Monte-Carlo simulations for a
total of $A\!=\!4000$ particles ($N\!=\!3200$, $Z\!=\!800$) have
been performed. Because of the many competing minima, a significantly 
larger system would take an unreasonably long time to thermalize on 
a modest work station (see details below).

The simulation volume for the above conditions consists of a cube of
length $L=73.7$ fm. While this value is larger than the electron
screening length $\lambda=26.6$~fm [see Eq.~(\ref{lambda})], it is
not sufficiently larger. Indeed, to minimize finite-size effects 
in a simulation with periodic boundary conditions one would like 
$\exp\bigl(-L/(2\lambda)\bigr)\!\ll\!1$. Clearly, this condition is 
not adequately satisfied. Therefore, in an effort to minimize the
contamination from finite-size effects, we reduce the electron 
screening length---arbitrarily---to the following value:
\begin{equation}
 \lambda \equiv 10~{\rm fm} \;.
\end{equation}
This value for $\lambda$ is adopted hereafter for all of our
simulations (see also~\cite{Mar98_PRC57}). This smaller screening 
length decreases slightly the Coulomb interaction at large distances,
which could promote the growth of slightly larger clusters. However, 
we do not expect this decrease in $\lambda$ to qualitatively change 
our results. We note that $\lambda$ is still larger than the typical
size of a heavy nucleus.

\begin{figure}[ht]

\begin{center}
\includegraphics[width=3in,angle=0,clip=false]{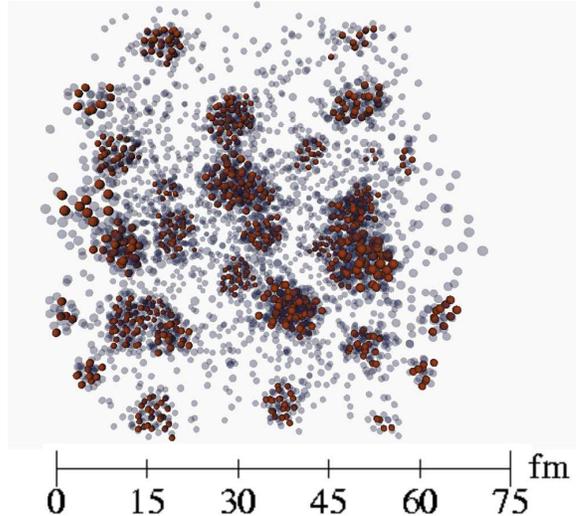}
\caption{(Color online) Monte-Carlo snapshot of a configuration of 
         $N\!=\!3200$ neutrons (light gray circles) and 
         $Z\!=\!800$ protons (dark red circles) at a 
         baryon density of $\rho\!=\!0.01~{\rm fm}^{-3}$,
	 a temperature of $T\!=\!1$ MeV, and an electron 
	 fraction of $Y_e\!=\!0.2$.  3D imaging courtesy of the FSU Visualization Laboratory.
} 
\label{Fig2}
\end{center}
\end{figure}


By far, the most time consuming part of the simulation is producing
suitable initial conditions. The simulations are started with the
$A\!=\!4000$ nucleons randomly distributed throughout the simulation
volume. Next, we perform a total of about 325,000 Metropolis sweeps starting at the higher temperature of $T\!=\!2$~MeV and reducing the temperature until
eventually reaching the target temperature of $T\!=\!1$~MeV, in a
``poor's-man'' attempt at simulated annealing.  Note that a Metropolis
sweep consists of a single trial move for each of the $A\!=\!4000$ particles in this system. We call this procedure cooking the pasta.

Results in this section are based on a statistical average of the final 50,000 sweeps. This yields a potential energy of $-5.385\pm 0.003$~MeV/A.  A sample configuration of the 4000 particles is shown in Fig.~\ref{Fig2}. The protons are strongly correlated into clusters (``nuclei'') as are a large number of neutrons. In addition, there is a low density neutron gas between the clusters. At this density it may be reasonable to think of the system as a high-density liquid of ``conventional'' nuclei immersed in a dilute neutron gas. A great virtue of the simulation is that one does not have to arbitrarily decide which nucleons cluster in nuclei and which ones remain in the gas. These ``decisions'' are being answered dynamically. Further, one can calculate modifications to nuclear properties due to the interactions. In a future work we plan to compare our simulation results to some conventional nuclear models.

Protons moving between clusters face a Coulomb barrier. This may 
inhibit the thermalization process and with it the formation of 
larger clusters. This could increase our results for $S({\bf q})$.
To test the thermalization of the pasta, our Metropolis Monte Carlo configuration was evolved further via molecular dynamics for a total time of 46,500 fm/c.  The molecular dynamics calculations will be described in future work.  This led to an increase in the peak of $S(q)$ by only about 10 percent.  Although these molecular dynamics results did not reveal a large secular change in the system with time, we caution that our cooking procedure may not have converged to the true thermal-equilibrium state.  The static structure factor $S(q)$ may still change with additional Metropolis Monte Carlo or molecular dynamics evolution.

\begin{figure}[ht]
\begin{center}
\vskip 0.3in

\includegraphics[height=3.3in,angle=-90,clip=false]{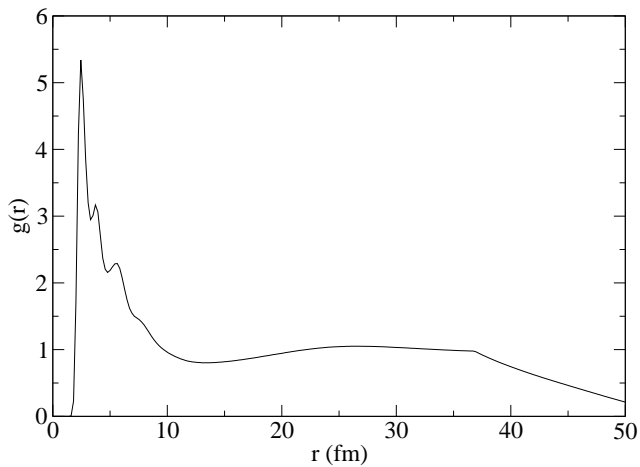}
\caption{Neutron-neutron correlation function at a 
         temperature of $T\!=\!1$ MeV, an electron fraction
	 of $Y_e\!=\!0.2$, and a baryon density of 
         $\rho\!=\!0.01~{\rm fm}^{-3}$.
	 These results show large finite-size effects 
	 beyond $r\!=\!L/2=36.9$~fm.}
\label{Fig3}

\end{center}
\end{figure}

The neutron-neutron correlation function $g(r)$ is displayed in
Fig.~\ref{Fig3}.  The two-neutron correlation function is calculated
by histograming the relative distances between neutrons.  The correlation function is very small at short distances because of the hard core in our $NN$ interaction. At intermediate distances $g(r)$ shows a large broad peak between $r\!=\!2$~fm and $r\!\simeq\!10$~fm. This corresponds to the other neutrons bound into a cluster. Superimposed on this broad peak one observes three (or four) sharper peaks corresponding to nearest, next-to-nearest, and next-to-next-to-nearest neighbors. These structures describe two-neutron correlations within the same cluster. At larger distances, between $10$ and $20$~fm, the correlation function shows a modest dip below one, suggesting that the attractive $NN$ interaction has shifted some neutrons from larger to smaller distances in order to form the clusters. Alternatively, Coulomb repulsion makes it less likely to
find two clusters separated by these radii. Finally, there is an
abrupt drop in $g(r)$ at a distance corresponding to half the value of
the simulation length ($r\!=\!L/2\!=\!36.9$~fm) caused by finite-size
effects.  We note that under our assumptions of periodic boundary
conditions, $g(r)$ is identically zero for $r\!>\!\sqrt{3}L/2$.

\begin{figure}[ht]
\begin{center}
\includegraphics[width=3in,angle=0,clip=false]{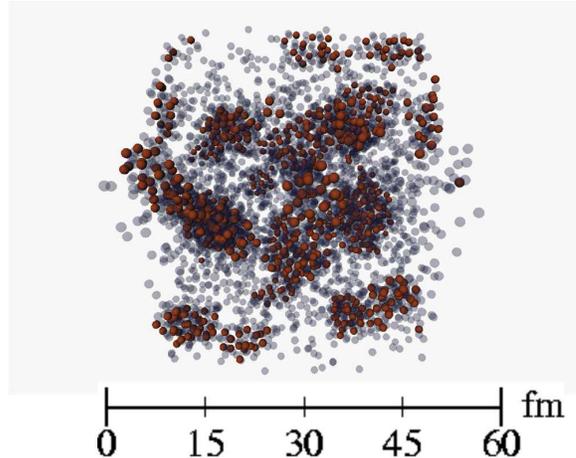}
\caption{(Color online) Monte-Carlo snapshot of a configuration of 
         $N\!=\!3200$ neutrons (light gray circles) and 
         $Z\!=\!800$ protons (dark red circles) at a 
         baryon density of $\rho\!=\!0.025~{\rm fm}^{-3}$,
	 a temperature of $T\!=\!1$ MeV, and an electron 
	 fraction of $Y_e\!=\!0.2$.  3D imaging courtesy of the FSU Visualization Laboratory.
} 
\label{Fig4}
\end{center}
\end{figure}

Increasing the density can change the nature of the pasta.  Figure ~\ref{Fig4}, shows a sample configuration of 4000 particles at a density of 0.025 fm$^{-3}$.
Note that although the density has increased, both the temperature and the
electron fraction have remained fixed at $T\!=\!1$~MeV and
$Y_e\!=\!0.2$, respectively.  At a density of $\rho\!=\!0.025~{\rm
fm}^{-3}$ (about $1/6$ of normal nuclear-matter saturation density)
the spherical clusters of Fig.~\ref{Fig2} start to coalesce into {\it
cylindrical-like} structures. The system appears to be segregated into 
high-density regions of cylindrical nuclei immersed in a dilute
neutron gas. As we continue to perform additional simulations, 
high-quality renderings of nucleon configurations for a variety of 
densities, temperatures, and electron fractions will be developed.

\begin{figure}[ht]
\begin{center}
\vskip 0.3in
\includegraphics[height=3.3in,angle=-90,clip=false]{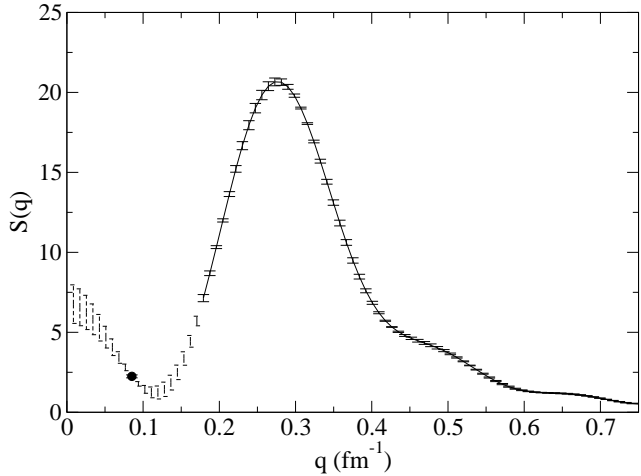}
\caption{Static structure factor $S({\bf q})$ versus momentum transfer 
         $q$ at a temperature of $T\!=\!1$ MeV, an electron fraction
	 of $Y_e\!=\!0.2$, and a baryon density of 
         $\rho\!=\!0.01~{\rm fm}^{-3}$.  The error bars are statistical only.  Finite size effects may be important at small $q$ as indicated by the dotted error bars.}
\label{Fig5}
\end{center}
\end{figure}

To compute the static structure factor $S(q)$, one numerically transforms the two-neutron correlation function, as indicated in Eq.~(\ref{sq4}). However, because of finite-size effects we truncate the integral at $r\!=\!L/2$ and assume $g(r)\!\equiv\!1$ for $r\!>\!L/2$. This procedure yields the results displayed in Fig.~\ref{Fig5}. There is a modest peak in $S(q\!=\!0)$ due to density fluctuations. Of course, the number of neutrons in our simulation remains fixed, yet fluctuations can take neutrons across the $r\!=\!L/2$ cutoff and these fluctuations will contribute to the value of $S(q)$ at $q\!=\!0$.


The error bars in Fig.~\ref{Fig5} are statistical only, based on the last 50,000 sweeps.  We caution that there may be finite-size effects 
at small momentum transfers.  The box size for our simulation at $\rho=0.01$ fm$^{-3}$ is $L\!=\!73.7 $~fm. This corresponds to a minimum momentum 
transfer of
\begin{equation}
 q_{\rm min} \approx \frac{2\pi}{L} = 0.085~{\rm fm^{-1}} \;.
 \label{qmin}
\end{equation}
Momentum transfers smaller than $q_{\rm min}$ correspond to
wavelengths larger than the simulation volume, so our results for
$q\!\alt 2q_{\rm min}$ may be sensitive to finite-size effects.
Indeed, for $q\!\alt 2q_{\rm min}$ the static structure factor was
observed to change significantly from one Metropolis run to the next.
To indicate the sensitivity of our results to finite-size effects,
Fig.~\ref{Fig5} displays the static structure factor in the
$q\!\le 2q_{\rm min}$ region with dotted error bars. Note that the point
$q\!=\!q_{\rm min}$ has been signaled out in Fig.~\ref{Fig5} to
indicate that it is more stable from one Metropolis run to the next,
because the weak vector-charge density [Eq.~(\ref{rhoq})] evaluated at
${\bf q}\!=\!q_{\rm min}\hat{\bf q}$ is invariant under a translation
of the system by a distance $L$ along $\hat{\bf q}$.

The static structure factor displays a large peak at
$q\!\approx\!0.3~{\rm fm}^{-1}$, corresponding to coherent scattering from many neutrons bound into a single cluster.  At smaller momentum transfers, $q\!\approx\!0.2~{\rm fm}^{-1}$, $S({\bf q})$ decreases because of ion screening. Here the neutrino wavelength is so long that it probes multiple clusters. These other clusters screen the weak charge and reduce the response.
At momentum transfers larger than $q\!\approx\!0.3~{\rm fm}^{-1}$,
the static structure factor decreases with increasing $q$. This is 
the effect of the cluster form factor. As the momentum transfer 
increases the neutrino can no longer scatter coherently from all 
the neutrons in a cluster because of the cluster's extended size.  
Thus, the observed peak in $S({\bf q})$ develops as a trade off 
between ion screening, which favors large $q$, and the cluster form 
factor, which favors small $q$.

In summary, one can divide the response of the pasta into the
following regions. At low-momentum transfers ($q\!\alt\!0.2~{\rm
fm}^{-1}$) the response is dominated by ion screening and density
fluctuations. For momentum transfers in the region
$q\!=\!0.2\!-\!0.4~{\rm fm}^{-1}$ one observes coherent scattering
from the pasta. At the larger momenta of $q\!=\!0.4\!-\!1~{\rm
fm}^{-1}$, the falling response reflects the pasta form
factor. Finally, the large momentum transfer region above $q\!=\!1~{\rm fm}^{-1}$ corresponds to quasielastic scattering from nearly free neutrons, as
$S(q\rightarrow\infty)\!=\!1$ [see Eq.~({\ref{slarge})].

\begin{figure}[ht]
\begin{center}
\vskip 0.3in
\includegraphics[height=3.3in,angle=-90,clip=false]{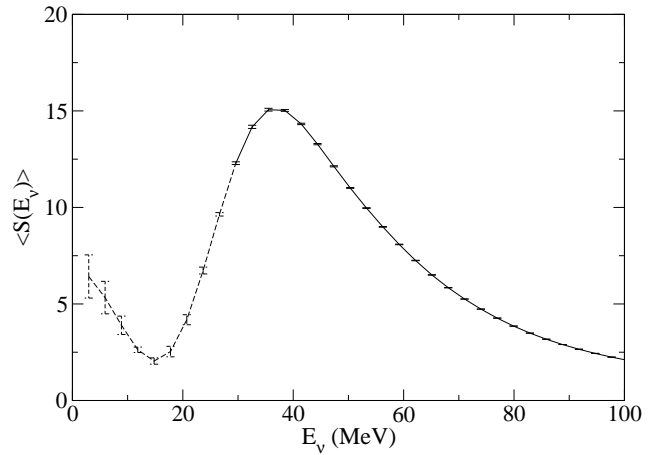}
\caption{Angle averaged static structure factor $\langle S(E_\nu) 
         \rangle$ versus neutrino energy $E_\nu$ at a temperature 
	 of $T\!=\!1$ MeV, an electron fraction of $Y_e\!=\!0.2$, 
	 and a baryon density of $\rho\!=\!0.01~{\rm fm}^{-3}$. The error bars are statistical only, see text.}
\label{Fig6}
\end{center}
\end{figure}

The angle averaged structure factor $\langle S(E_\nu) \rangle$ 
[defined in Eq.~(\ref{save})] is shown in Fig.~\ref{Fig6}. Note
that the integral in Eq.~(\ref{save2}) was also truncated at
$r\!=\!L/2$ because of finite-size effects. The averaged 
structure factor shows a broad peak for neutrino energies near 
$40$~MeV. Indeed, the transport cross section is significantly 
enhanced by coherence effects for neutrino energies from about 
$20$ to $80$~MeV. The impact of this coherence on the neutrino
mean-free path [Eq.~(\ref{mfp})] is displayed in Fig.~\ref{Fig7}. 
Also shown in the figure is the mean-free path obtained by 
ignoring coherence effects by setting 
$\langle S(E_\nu) \rangle\!=\!1$ in Eq.~(\ref{mfp}). Coherence
significantly reduces the mean-free path for neutrino energies 
in the range $E_\nu\!=\!15\!-\!120$~MeV.  Again, finite size effects may be important for low neutrino energies.

\begin{figure}[ht]
\begin{center}
\vskip 0.3in

\includegraphics[height=3.3in,angle=-90,clip=false]{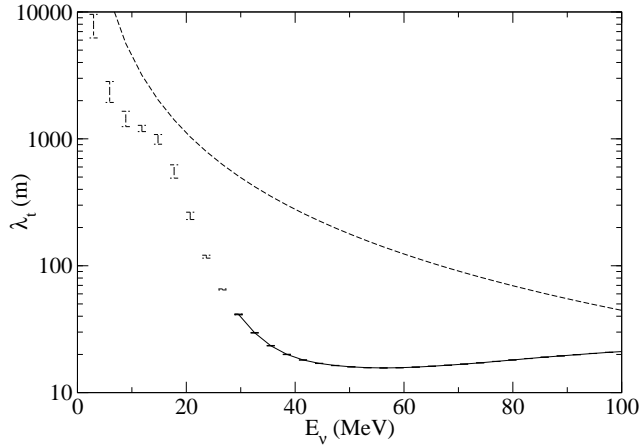}
\caption{Transport mean-free path $\lambda_t$ for $\nu_\mu$ or 
         $\nu_\tau$ versus neutrino energy $E_\nu$ at a 
         baryon density of $\rho\!=\!0.01~{\rm fm}^{-3}$,
	 a temperature of $T\!=\!1$ MeV, and an electron 
	 fraction of $Y_e\!=\!0.2$. The solid line and error bars include 
         full coherence effects while the dashed line is 
         obtained by using $\langle S(E) \rangle\!=\!1$ in 
         Eq.~(\ref{mfp}).  Error bars are statistical only.  Finite size effects may be important for low $E_\nu$ as indicated by the dotted error bars.}
\label{Fig7}
\end{center}
\end{figure}

\section{Conclusions}
\label{conclusions}
Neutron-rich matter is expected to have a complex structure at
subnuclear densities. Complex pasta phases may result from 
frustration through the competition between an attractive
nuclear interaction and the Coulomb repulsion. Neutrino interactions 
with the pasta may be important for properties of core-collapse 
supernovae, such as its electron fraction.

In this work we have employed a semi-classical model to simulate the
dynamics of the pasta phase of neutron-rich matter. Although our model
is very simple, it nonetheless retains the crucial physics of
frustration. Using a Metropolis Monte-Carlo algorithm, the partition
function was computed for a system of 4000 nucleons at a given
temperature and density. We find that almost all protons and most of
the neutrons cluster into ``nuclei'' that are surrounded by a dilute
neutron gas.

Observables computed in our simulations included the neutron-neutron
correlation function. This calculation was implemented by constructing
a histogram of all relative neutron distances. The two-neutron
correlation function $g(r)$ gives the probability of finding a neutron
at a distance $r$ away from a reference neutron. A large peak in
$g(r)$ at intermediate distances ($r\!=\!2\!-\!10$~fm) is found. This
reflects the presence of the other neutrons in the cluster. 

The static structure factor $S({\bf q})$, a fundamental observable
obtained from the Fourier transform of the two-neutron correlation
function, describes the degree of coherence for neutrino-nucleon
elastic scattering. For small momentum transfers, $S({\bf q})$
describes density fluctuations and ion screening. In this region
the neutrino wavelength is longer than the average inter-cluster
separation, thereby allowing other clusters to screen the weak
charge of a given cluster. At momentum transfers of approximately 
$q\!=\!0.2\!-\!0.4~{\rm fm}^{-1}$, the static structure factor
develops a large peak, associated to the coherent scattering from 
all the neutrons in the cluster. This coherence is responsible for a
significant reduction in the neutrino mean-free path. To our
knowledge, these represent the first consistent calculation of the 
neutrino mean-free path in nonuniform neutron-rich matter.

However, much remains to be done. First, one needs to focus on the
thermalization of our simulations. It is difficult to insure that the
system has reached thermal equilibrium because the Coulomb barrier hinders the motion of individual protons. Second, one must further investigate the impact of finite-size effects and the simple treatment of long-range Coulomb interactions on our simulations. This may require simulations with larger numbers of particles, as it is difficult to fit a long-wavelength neutrino into the present
simulation volume. Third, while we have focused here on the vector
part of the weak-neutral-current response, because it can be greatly
enhanced by coherence, one should extend the study to the axial-vector
(or spin) response, as it dominates the scattering when it is coming
from uncorrelated nucleons. Further, one should also calculate
charged-current interactions in nonuniform matter.  Finally, in the
present contribution no effort was made to optimize the $NN$
interaction. While it may be advantageous to do so, any {\it
accurately calibrated} interaction must retain the essential features
of frustration. Moreover, more sophisticated interactions that include
momentum and/or density dependence will significantly increase the
computational demands. At present, we are checking our results against
more sophisticated simulations using molecular dynamics, studying finite size effects in larger simulations, exploring the temperature and density dependence of our results, and calculating the dynamical response. These results
will be presented in a future contribution~\cite{Hor_FUTURE}.

\begin{acknowledgments}

We acknowledge useful discussions with Wolfgang Bauer, Adam Burrows, 
Vladimir Dobrosavljevic, Thomas Janka, Sanjay Reddy, Pedro Schlottmann, Romualdo de Souza, and Victor Viola. C.J.H. thanks the Aspen Center for Physics, the Max Plank Institute for Astrophysics, and the Institute for Nuclear Theory at the University of Washington for their hospitality. M.A.P.G. acknowledges partial support from Indiana University and FICYT. J.P. thanks David Banks and the staff at the FSU Visualization Laboratory for their help.  This work was supported in part by DOE grants DE-FG02-87ER40365 and DE-FG05-92ER40750.

\end{acknowledgments}

\vfill\eject

\bibliography{ReferencesJP}

\begin{thebibliography}{34}
\expandafter\ifx\csname natexlab\endcsname\relax\def\natexlab#1{#1}\fi
\expandafter\ifx\csname bibnamefont\endcsname\relax
  \def\bibnamefont#1{#1}\fi
\expandafter\ifx\csname bibfnamefont\endcsname\relax
  \def\bibfnamefont#1{#1}\fi
\expandafter\ifx\csname citenamefont\endcsname\relax
  \def\citenamefont#1{#1}\fi
\expandafter\ifx\csname url\endcsname\relax
  \def\url#1{\texttt{#1}}\fi
\expandafter\ifx\csname urlprefix\endcsname\relax\def\urlprefix{URL }\fi
\providecommand{\bibinfo}[2]{#2}
\providecommand{\eprint}[2][]{\url{#2}}

\bibitem[{\citenamefont{Ravenhall et~al.}(1983)\citenamefont{Ravenhall,
  Pethick, and Wilson}}]{Rav83_PRL50}
\bibinfo{author}{\bibfnamefont{D.~G.} \bibnamefont{Ravenhall}},
  \bibinfo{author}{\bibfnamefont{C.~J.} \bibnamefont{Pethick}},
  \bibnamefont{and} \bibinfo{author}{\bibfnamefont{J.~R.}
  \bibnamefont{Wilson}}, \bibinfo{journal}{Phys. Rev. Lett.}
  \textbf{\bibinfo{volume}{50}}, \bibinfo{pages}{2066} (\bibinfo{year}{1983}).

\bibitem[{\citenamefont{Hashimoto et~al.}(1984)\citenamefont{Hashimoto, Seki,
  and Yamada}}]{Has84_PTP71}
\bibinfo{author}{\bibfnamefont{M.}~\bibnamefont{Hashimoto}},
  \bibinfo{author}{\bibfnamefont{H.}~\bibnamefont{Seki}}, \bibnamefont{and}
  \bibinfo{author}{\bibfnamefont{M.}~\bibnamefont{Yamada}},
  \bibinfo{journal}{Prog. Theor. Phys.} \textbf{\bibinfo{volume}{71}},
  \bibinfo{pages}{320} (\bibinfo{year}{1984}).

\bibitem[{\citenamefont{Williams and Koonin}(1985)}]{Wil85_NPA435}
\bibinfo{author}{\bibfnamefont{R.}~\bibnamefont{Williams}} \bibnamefont{and}
  \bibinfo{author}{\bibfnamefont{S.~E.} \bibnamefont{Koonin}},
  \bibinfo{journal}{Nucl. Phys.} \textbf{\bibinfo{volume}{A435}},
  \bibinfo{pages}{844} (\bibinfo{year}{1985}).

\bibitem[{\citenamefont{Oyamatsu}(1993)}]{Oya93_NPA561}
\bibinfo{author}{\bibfnamefont{K.}~\bibnamefont{Oyamatsu}},
  \bibinfo{journal}{Nucl. Phys} \textbf{\bibinfo{volume}{A561}},
  \bibinfo{pages}{431} (\bibinfo{year}{1993}).

\bibitem[{\citenamefont{Sumiyoshi et~al.}(1995)\citenamefont{Sumiyoshi,
  Oyamatsu, and Toki}}]{Sum95_NPA595}
\bibinfo{author}{\bibfnamefont{K.}~\bibnamefont{Sumiyoshi}},
  \bibinfo{author}{\bibfnamefont{K.}~\bibnamefont{Oyamatsu}}, \bibnamefont{and}
  \bibinfo{author}{\bibfnamefont{H.}~\bibnamefont{Toki}},
  \bibinfo{journal}{Nucl. Phys} \textbf{\bibinfo{volume}{A595}},
  \bibinfo{pages}{327} (\bibinfo{year}{1995}).

\bibitem[{\citenamefont{Lorenz et~al.}(1993)\citenamefont{Lorenz, Ravenhall,
  and Pethick}}]{Lor93_PRL70}
\bibinfo{author}{\bibfnamefont{C.~P.} \bibnamefont{Lorenz}},
  \bibinfo{author}{\bibfnamefont{D.~G.} \bibnamefont{Ravenhall}},
  \bibnamefont{and} \bibinfo{author}{\bibfnamefont{C.~J.}
  \bibnamefont{Pethick}}, \bibinfo{journal}{Phys. Rev. Lett.}
  \textbf{\bibinfo{volume}{70}}, \bibinfo{pages}{379} (\bibinfo{year}{1993}).

\bibitem[{\citenamefont{Toulouse}(1977)}]{Tou77_CP2}
\bibinfo{author}{\bibfnamefont{G.}~\bibnamefont{Toulouse}},
  \bibinfo{journal}{Commun. Phys.} \textbf{\bibinfo{volume}{2}},
  \bibinfo{pages}{115} (\bibinfo{year}{1977}).

\bibitem[{\citenamefont{Villain}(1977)}]{Vil77_JP10}
\bibinfo{author}{\bibfnamefont{J.}~\bibnamefont{Villain}},
  \bibinfo{journal}{Jour. Phys. C} \textbf{\bibinfo{volume}{10}},
  \bibinfo{pages}{1717} (\bibinfo{year}{1977}).

\bibitem[{\citenamefont{Liebmann}(1986)}]{Lie86_SV251}
\bibinfo{author}{\bibfnamefont{R.}~\bibnamefont{Liebmann}},
  \emph{\bibinfo{title}{Statistical Mechanics of Periodic Frustrated Ising
  Systems}}, vol. \bibinfo{volume}{251} (\bibinfo{publisher}{Springer-Verlag},
  \bibinfo{address}{Berlin Heidelberg}, \bibinfo{year}{1986}).

\bibitem[{\citenamefont{Diep}(1994)}]{Die94_WS}
\bibinfo{author}{\bibfnamefont{H.~T.} \bibnamefont{Diep}},
  \emph{\bibinfo{title}{Magnetic Systems with Competing Interactions}}
  (\bibinfo{publisher}{World Scientific}, \bibinfo{address}{Singapore},
  \bibinfo{year}{1994}).

\bibitem[{\citenamefont{Camacho}(1996)}]{Cam96_PRL77}
\bibinfo{author}{\bibfnamefont{C.~J.} \bibnamefont{Camacho}},
  \bibinfo{journal}{Phys. Rev. Lett.} \textbf{\bibinfo{volume}{77}},
  \bibinfo{pages}{2324} (\bibinfo{year}{1996}).

\bibitem[{\citenamefont{Wannier}(1950)}]{Wan50_PR79}
\bibinfo{author}{\bibfnamefont{G.~H.} \bibnamefont{Wannier}},
  \bibinfo{journal}{Phys. Rev.} \textbf{\bibinfo{volume}{79}},
  \bibinfo{pages}{357} (\bibinfo{year}{1950}).

\bibitem[{\citenamefont{Kirkpatrick and Sherrington}(1978)}]{Kir78_PRB17}
\bibinfo{author}{\bibfnamefont{S.}~\bibnamefont{Kirkpatrick}} \bibnamefont{and}
  \bibinfo{author}{\bibfnamefont{D.}~\bibnamefont{Sherrington}},
  \bibinfo{journal}{Phys. Rev. B} \textbf{\bibinfo{volume}{17}},
  \bibinfo{pages}{4384} (\bibinfo{year}{1978}).

\bibitem[{\citenamefont{Papadimitriou and Kenneth~Steiglitz}(1998)}]{Pap98_Dov}
\bibinfo{author}{\bibfnamefont{C.~H.} \bibnamefont{Papadimitriou}}
  \bibnamefont{and}
  \bibinfo{author}{\bibfnamefont{K.}~\bibnamefont{Kenneth~Steiglitz}},
  \emph{\bibinfo{title}{Combinatorial Optimization: Algorithms and Complexity}}
  (\bibinfo{publisher}{Dover}, \bibinfo{address}{Mineola, New York},
  \bibinfo{year}{1998}).

\bibitem[{\citenamefont{Bodmer et~al.}(1980)\citenamefont{Bodmer, Panos, and
  MacKellar}}]{Bod80_PRC22}
\bibinfo{author}{\bibfnamefont{A.~R.} \bibnamefont{Bodmer}},
  \bibinfo{author}{\bibfnamefont{C.~N.} \bibnamefont{Panos}}, \bibnamefont{and}
  \bibinfo{author}{\bibfnamefont{A.~D.} \bibnamefont{MacKellar}},
  \bibinfo{journal}{Phys. Rev. C} \textbf{\bibinfo{volume}{22}},
  \bibinfo{pages}{1025} (\bibinfo{year}{1980}).

\bibitem[{\citenamefont{Schlagel and Pandharipande}(1987)}]{Sch87_PRC36}
\bibinfo{author}{\bibfnamefont{T.~J.} \bibnamefont{Schlagel}} \bibnamefont{and}
  \bibinfo{author}{\bibfnamefont{V.~R.} \bibnamefont{Pandharipande}},
  \bibinfo{journal}{Phys. Rev. C} \textbf{\bibinfo{volume}{36}},
  \bibinfo{pages}{162} (\bibinfo{year}{1987}).

\bibitem[{\citenamefont{Peilert et~al.}(1991)\citenamefont{Peilert, Randrup,
  Stocker, and Greiner}}]{Pei91_PLB260}
\bibinfo{author}{\bibfnamefont{G.}~\bibnamefont{Peilert}},
  \bibinfo{author}{\bibfnamefont{J.}~\bibnamefont{Randrup}},
  \bibinfo{author}{\bibfnamefont{H.}~\bibnamefont{Stocker}}, \bibnamefont{and}
  \bibinfo{author}{\bibfnamefont{W.}~\bibnamefont{Greiner}},
  \bibinfo{journal}{Phys. Lett. B} \textbf{\bibinfo{volume}{260}},
  \bibinfo{pages}{271} (\bibinfo{year}{1991}).

\bibitem[{\citenamefont{Maruyama et~al.}(1998)\citenamefont{Maruyama, Niita,
  Oyamatsu, Maruyama, Chiba, and Iwamoto}}]{Mar98_PRC57}
\bibinfo{author}{\bibfnamefont{T.}~\bibnamefont{Maruyama}},
  \bibinfo{author}{\bibfnamefont{K.}~\bibnamefont{Niita}},
  \bibinfo{author}{\bibfnamefont{K.}~\bibnamefont{Oyamatsu}},
  \bibinfo{author}{\bibfnamefont{T.}~\bibnamefont{Maruyama}},
  \bibinfo{author}{\bibfnamefont{S.}~\bibnamefont{Chiba}}, \bibnamefont{and}
  \bibinfo{author}{\bibfnamefont{A.}~\bibnamefont{Iwamoto}},
  \bibinfo{journal}{Phys. Rev. C} \textbf{\bibinfo{volume}{57}},
  \bibinfo{pages}{655} (\bibinfo{year}{1998}).

\bibitem[{\citenamefont{Watanabe
  et~al.}(2003{\natexlab{a}})\citenamefont{Watanabe, Sato, Yasuoka, and
  Ebisuzaki}}]{Wat03_PRC68}
\bibinfo{author}{\bibfnamefont{G.}~\bibnamefont{Watanabe}},
  \bibinfo{author}{\bibfnamefont{K.}~\bibnamefont{Sato}},
  \bibinfo{author}{\bibfnamefont{K.}~\bibnamefont{Yasuoka}}, \bibnamefont{and}
  \bibinfo{author}{\bibfnamefont{T.}~\bibnamefont{Ebisuzaki}},
  \bibinfo{journal}{Phys. Rev. C} \textbf{\bibinfo{volume}{68}},
  \bibinfo{pages}{035806} (\bibinfo{year}{2003}{\natexlab{a}}),
  \eprint[http://arXiv.org/abs]{nucl-th/0308007}.

\bibitem[{\citenamefont{Watanabe
  et~al.}(2003{\natexlab{b}})\citenamefont{Watanabe, Sato, Yasuoka, and
  Ebisuzaki}}]{Wat03_xxx}
\bibinfo{author}{\bibfnamefont{G.}~\bibnamefont{Watanabe}},
  \bibinfo{author}{\bibfnamefont{K.}~\bibnamefont{Sato}},
  \bibinfo{author}{\bibfnamefont{K.}~\bibnamefont{Yasuoka}}, \bibnamefont{and}
  \bibinfo{author}{\bibfnamefont{T.}~\bibnamefont{Ebisuzaki}}
  (\bibinfo{year}{2003}{\natexlab{b}}),
  \eprint[http://arXiv.org/abs]{nucl-th/0311083}.

\bibitem[{\citenamefont{Horowitz}(1997)}]{Hor97_PRD55}
\bibinfo{author}{\bibfnamefont{C.~J.} \bibnamefont{Horowitz}},
  \bibinfo{journal}{Phys. Rev. D} \textbf{\bibinfo{volume}{55}},
  \bibinfo{pages}{4577} (\bibinfo{year}{1997}).

\bibitem[{\citenamefont{Igashira et~al.}(1986)\citenamefont{Igashira, Kitazawa,
  Komano, and Yamamuro}}]{Iga86_NPA457}
\bibinfo{author}{\bibfnamefont{M.}~\bibnamefont{Igashira}},
  \bibinfo{author}{\bibfnamefont{H.}~\bibnamefont{Kitazawa}},
  \bibinfo{author}{\bibfnamefont{H.}~\bibnamefont{Komano}}, \bibnamefont{and}
  \bibinfo{author}{\bibfnamefont{N.}~\bibnamefont{Yamamuro}},
  \bibinfo{journal}{Nucl. Phys.} \textbf{\bibinfo{volume}{A457}},
  \bibinfo{pages}{301} (\bibinfo{year}{1986}).

\bibitem[{\citenamefont{Vretenar et~al.}(2001)\citenamefont{Vretenar, Paar,
  Ring, and Lalazissis}}]{Vre01_PRC63}
\bibinfo{author}{\bibfnamefont{D.}~\bibnamefont{Vretenar}},
  \bibinfo{author}{\bibfnamefont{N.}~\bibnamefont{Paar}},
  \bibinfo{author}{\bibfnamefont{P.}~\bibnamefont{Ring}}, \bibnamefont{and}
  \bibinfo{author}{\bibfnamefont{G.~A.} \bibnamefont{Lalazissis}},
  \bibinfo{journal}{Phys. Rev. C} \textbf{\bibinfo{volume}{63}},
  \bibinfo{pages}{047301} (\bibinfo{year}{2001}),
  \eprint[http://arXiv.org/abs]{nucl-th/0009057}.

\bibitem[{\citenamefont{Reddy et~al.}(2000)\citenamefont{Reddy, Bertsch, and
  Prakash}}]{Red00_PLB475}
\bibinfo{author}{\bibfnamefont{S.}~\bibnamefont{Reddy}},
  \bibinfo{author}{\bibfnamefont{G.}~\bibnamefont{Bertsch}}, \bibnamefont{and}
  \bibinfo{author}{\bibfnamefont{M.~P.} \bibnamefont{Prakash}},
  \bibinfo{journal}{Phys. Lett.} \textbf{\bibinfo{volume}{B475}},
  \bibinfo{pages}{1} (\bibinfo{year}{2000}).

\bibitem[{\citenamefont{Lattimer and Swesty}(1992)}]{Lat92_NPA535}
\bibinfo{author}{\bibfnamefont{J.~M.} \bibnamefont{Lattimer}} \bibnamefont{and}
  \bibinfo{author}{\bibfnamefont{F.~D.} \bibnamefont{Swesty}},
  \bibinfo{journal}{Nucl. Phys.} \textbf{\bibinfo{volume}{A535}},
  \bibinfo{pages}{331} (\bibinfo{year}{1992}).

\bibitem[{\citenamefont{Liebendoerfer et~al.}(2002)\citenamefont{Liebendoerfer,
  Messer, Mezzacappa, Bruenn, Cardall, and Thielemann}}]{Lie02_APJ}
\bibinfo{author}{\bibfnamefont{M.}~\bibnamefont{Liebendoerfer}},
  \bibinfo{author}{\bibfnamefont{O.~E.~B.} \bibnamefont{Messer}},
  \bibinfo{author}{\bibfnamefont{A.}~\bibnamefont{Mezzacappa}},
  \bibinfo{author}{\bibfnamefont{S.~W.} \bibnamefont{Bruenn}},
  \bibinfo{author}{\bibfnamefont{C.~Y.} \bibnamefont{Cardall}},
  \bibnamefont{and} \bibinfo{author}{\bibfnamefont{F.~K.}
  \bibnamefont{Thielemann}} (\bibinfo{year}{2002}),
  \eprint[http://arXiv.org/abs]{astro-ph/0207036}.

\bibitem[{\citenamefont{Jeschonnek and Van~Orden}(2002)}]{Jes02_PRD65}
\bibinfo{author}{\bibfnamefont{S.}~\bibnamefont{Jeschonnek}} \bibnamefont{and}
  \bibinfo{author}{\bibfnamefont{J.~W.} \bibnamefont{Van~Orden}},
  \bibinfo{journal}{Phys. Rev. D} \textbf{\bibinfo{volume}{65}},
  \bibinfo{pages}{094038} (\bibinfo{year}{2002}), \bibinfo{note}{and references
  therein}.

\bibitem[{\citenamefont{Fetter and Walecka}(1971)}]{Fet71_MH}
\bibinfo{author}{\bibfnamefont{A.~L.} \bibnamefont{Fetter}} \bibnamefont{and}
  \bibinfo{author}{\bibfnamefont{J.~D.} \bibnamefont{Walecka}},
  \emph{\bibinfo{title}{Quantum Theory of Many-Particle Systems}}
  (\bibinfo{publisher}{McGraw-Hill}, \bibinfo{address}{New York},
  \bibinfo{year}{1971}).

\bibitem[{\citenamefont{Chin}(1977)}]{Chi77_AOP108}
\bibinfo{author}{\bibfnamefont{S.~A.} \bibnamefont{Chin}},
  \bibinfo{journal}{Ann. of Phys.} \textbf{\bibinfo{volume}{108}},
  \bibinfo{pages}{301} (\bibinfo{year}{1977}).

\bibitem[{\citenamefont{Metropolis et~al.}(1953)\citenamefont{Metropolis,
  Rosenbluth, Rosenbluth, Teller, and Teller}}]{Met53_JCP21}
\bibinfo{author}{\bibfnamefont{N.}~\bibnamefont{Metropolis}},
  \bibinfo{author}{\bibfnamefont{A.}~\bibnamefont{Rosenbluth}},
  \bibinfo{author}{\bibfnamefont{M.}~\bibnamefont{Rosenbluth}},
  \bibinfo{author}{\bibfnamefont{A.}~\bibnamefont{Teller}}, \bibnamefont{and}
  \bibinfo{author}{\bibfnamefont{E.}~\bibnamefont{Teller}},
  \bibinfo{journal}{Jour. Chem Phys.} \textbf{\bibinfo{volume}{21}},
  \bibinfo{pages}{1087} (\bibinfo{year}{1953}).

\bibitem[{\citenamefont{Horowitz}(2002)}]{Hor02_PRD65}
\bibinfo{author}{\bibfnamefont{C.~J.} \bibnamefont{Horowitz}},
  \bibinfo{journal}{Phys. Rev. D} \textbf{\bibinfo{volume}{65}},
  \bibinfo{pages}{043001} (\bibinfo{year}{2002}).

\bibitem[{\citenamefont{Freedman et~al.}(1977)\citenamefont{Freedman, Schramm,
  and Tubbs}}]{Fre77_ARNS27}
\bibinfo{author}{\bibfnamefont{D.~Z.} \bibnamefont{Freedman}},
  \bibinfo{author}{\bibfnamefont{D.~N.} \bibnamefont{Schramm}},
  \bibnamefont{and} \bibinfo{author}{\bibfnamefont{D.~L.} \bibnamefont{Tubbs}},
  \bibinfo{journal}{Annu. Rev. Nucl. Sci.} \textbf{\bibinfo{volume}{27}},
  \bibinfo{pages}{167} (\bibinfo{year}{1977}).

\bibitem[{\citenamefont{Egelstaff}(1967)}]{liquid}
\bibinfo{author}{\bibfnamefont{P.~A.} \bibnamefont{Egelstaff}},
  \emph{\bibinfo{title}{An Introduction to the Liquid State}}
  (\bibinfo{publisher}{Academic Press}, \bibinfo{address}{New York},
  \bibinfo{year}{1967}).

\bibitem[{\citenamefont{Horowitz et~al.}()\citenamefont{Horowitz,
  P\'{e}rez-Garc\'{\i }a, and Piekarewicz}}]{Hor_FUTURE}
\bibinfo{author}{\bibfnamefont{C.~J.} \bibnamefont{Horowitz}},
  \bibinfo{author}{\bibfnamefont{M.~A.} \bibnamefont{P\'{e}rez-Garc\'{\i }a}},
  \bibnamefont{and}
  \bibinfo{author}{\bibfnamefont{J.}~\bibnamefont{Piekarewicz}},
  \bibinfo{note}{in preparation}.

\end{thebibliography}
\end{document}